# Role of charge compensation mechanism on phase formation, dielectric and ferroelectric properties in aliovalent $Gd^{3+}$ ion modified $PbMg_{1/3}Nb_{2/3}O_3$ ceramics

Adityanarayan H. Pandey* and Surya Mohan Gupta*

*Homi Bhabha National Institute, Anushaktinagar, Mumbai-400094, India.*
*Laser Materials Section, Raja Ramanna Centre for Advanced Technology, Indore-452013, India.*

*Email – anbp.phy@gmail.com (A. H. Pandey), surya@rrcat.gov.in (S. M. Gupta)

## Abstract

Phase, microstructural, dielectric and ferroelectric properties investigation clearly reveal that the charge imbalance created due to alio-valent $Gd^{3+}$-ion substitution at $Pb^{2+}$ site in lead magnesium niobate should be compensated by changing B-site cation ratio instead of creating A-site or B-site vacancies. Microstructure analysis along with elemental mapping exhibit segregation of magnesium oxide (MgO) and gadolinium niobate ($GdNbO_4$) phase, which is observed to remain invariant to the charge compensation mechanism. Fitting of frequency dependent temperature of dielectric constant maximum $T_m$ (temperature of $\varepsilon_{max}$) and the Mydosh parameter "$K$" calculated from the $\varepsilon_{max}$ (T) clearly reveals critical slowing down of polar nano-regions (PNRs) dynamics leading to super-dipolar glass state in $Gd^{3+}$-ion substituted PMN, when the charge imbalance is compensated by creating B-site vacancy or changing B-site cations ratio. Precipitation of secondary (pyrochlore) phase is resulted in $Gd^{3+}$-ion substituted PMN when the charge imbalance is compensated by creating A-site vacancies, which causes reduction in interaction among the PNRs and circumventing the super-dipolar glass state.

## 1. Introduction

Relaxor ferroelectrics (RFEs) are fascinating because of their (i) unusual properties such as high and frequency dependent dielectric constants, minimal hysteresis loss, and large electromechanical strains that are not yet understood and (ii) utility in multilayer capacitors, actuators, transducers, and sensors etc. from technological perspective [1-5]. The RFEs, particularly, lead magnesium niobate (PMN), are described by nano sized cation and structural disorder, which leads to random electric/strain/bond fields and local phase fluctuations [3,4]. The random electric field is caused by the presence of chemically ordered regions (CORs) and polar nano regions (PNRs), which are believed to be responsible for the high dielectric and distinctive piezoelectric properties [6,7]. The CORs are Mg/Nb superstructure regions having Fm3m crystal structure as described by "random layer model" and appear below ~1000 K [8,9]. The PNRs are parmanently correlated dipole moment regions with a rhombohedral R3m crystal symmetry appear below the Burns temperature $T_B$ ~ 600 K [6,7,10]. Local characterization tools such as Raman spectroscopy, diffused neutron scattering and transmission electron microscopy (TEM) shows that number/size of the CORs remains unchanged in the temperature range of 5-800 K [11-13]. However, number/size of the PNRs grow on cooling below the $T_B$ and these regions freeze to ferroelectric nano sized domain state below a temperature $T_f$ ~220 K, which is also termed as ergodic to nonergodic transformation state akin to glasses [6,7,10,14]. At present, there are two acceptable models, (i) dipole glass model [6,15,16] and (ii) random-field model [17,18] for describing unique dielectric properties and the formation of nano-polar state in the PMN. However, origin of the CORs and inability of the PNRs to grow in micron size below the $T_m$ in zero-field cooling condition is not clear.

The CORs and PNRs of PMN are correlated and reported to be influenced by application of an external electric field and/or A/B-site chemical substitution by alio-valent ions [19-26]. Application of an external electric field has revealed a phase transition from glass phase to ferroelectric phase at a critical field of ~2 kV/cm in the temperature range of 160 K - 200 K [19]. On the other hand, B-site modification of PMN with Ti and Zr leads to reduction in size of the CORs and enhancement in size of the PNRs from nano to micron size i.e., relaxor to ferroelectric phase transition [20-23]. The A-site modification of PMN by La, Sm, Nd, Tb, Pr, etc. causes enhancement in size of the CORs and reduction in size of the PNRs leading to increase relaxor-like dielectric characteristics [22,24-27].

Alio-valent $La^{3+}$-ion substitution at the $Pb^{2+}$ site in PZT is extensively studied for electro-optic application and dielectric properties of the PZT is significantly modified [28]. The charge imbalance due to $La^{3+}$ ion at the $Pb^{2+}$ ion site is already reported to compensate by creating A or B-site vacancies [29]. The maximum dielectric constant ($\varepsilon_{max}$) and corresponding $T_m$ (temperature of $\varepsilon_{max}$) has been shown to depend upon the charge imbalance compensation mechanism. Similarly, charge imbalance compensation mechanism plays an important role in single perovskite phase formation in La-ion substituted morphotropic phase composition of PMN-PT 65/35 [30]. The aim of the present work is to correlate phase, microstructure, dielectric and ferroelectric properties with the charge compensation mechanism for magnetic rare-earth ion ($Gd^{3+}$) substituted PMN. The charge imbalance due to $Gd^{3+}$ at the $Pb^{2+}$ site of PMN is compensated by (i) creating A-vacancies ($Pb_{1-2x/3}Gd_x(Mg_{1/3}Nb_{2/3})O_3$) (ii) B-site vacancies ($Pb_{1-x}Gd_x(Mg_{1/3}Nb_{2/3})_{1-x/4}O_3$) and (iii) changing the B-site cations ratio ($Pb_{1-x}Gd_x(Mg_{1+x/3}Nb_{2-x/3})O_3$) and correlations are established.

## 2. Experimental

All ceramic samples of Gd-ion (x = 0.04) modified PMN [where charge imbalance is compensated by (i) creating Pb-ion vacancies ($Pb_{1-2x/3}Gd_x(Mg_{1/3}Nb_{2/3})O_3$; designated as PGMN4-VA) or (ii) creating Mg/Nb-ions vacancies ($Pb_{1-x}Gd_x(Mg_{1/3}Nb_{2/3})_{1-x/4}O_3$; designated as PGMN4-VB) or (iii) changing the ratio of Mg/Nb-ions ($Pb_{1-x}Gd_x(Mg_{1+x/3}Nb_{2-x/3})O_3$; designated as PGMN4-R) of the $ABO_3$ structure] are synthesized by the Columbite precursor method using high purity chemicals ('Alfa aesar' make): PbO (99.9%), $Gd_2O_3$ (99.99%), MgO (99.95%), $Nb_2O_5$ (99.9%). The Columbite precursor, magnesium niobate, $MgNb_2O_6$ for PGMN4-VA, PGMN4-VB and ($Mg_{(1+x)/3}Nb_{(2-x)/3}O_{4-x/2}$, for $x$ = 0.04) for PGMN4-R are prepared by mixing pre-determined amounts of MgO and $Nb_2O_5$ in ethanol and ball mill in zirconia bowl for 24 hours using zirconia grinding media. The slurry is dried in oven at 70 $^oC$ and the dried powder is calcined at 1200 $^oC$ for 2 hours. The calcined powder is once again ball milled for 24 hours and then dried similarly as mentioned above. Phase formation of the Columbite precursor is confirmed by using x-ray diffraction (XRD). The Columbite precursor is then mixed and ball milled with the predetermined amounts of PbO and $Gd_2O_3$ powders. Details about the ceramic fabrication are similar to the earlier report [26, 31]. In all the sintered ceramics, a sintered density greater than 97% of the theoretical is achieved. The calcined powder and the sintered specimens are analysed for phase using the XRD pattern. The XRD data is recorded on a Philips X'Pert diffractometer (Cu $K_\alpha$ source with $\lambda$ = 1.54 Å) using a scan rate of 0.5°/min with 0.01° step size in the $2\theta$ range 15°–90° to confirm the crystal structure and the phase. Rietveld refinement of the XRD pattern is carried out using Fullprof software to refine the lattice parameters of the crystal structure [32]. Field Emission Scanning Electron Microscope (FE-SEM, Carl Zeiss, SIGMA) equipped with energy dispersive spectroscopy (Oxford Inca X-Act LN2 free) is used for grain size, morphology and elemental determination from a fractured surface. The fractured surfaces are sputtered by gold. The linear intercept method (Image J software) has been used to measure the average grain size.

The sintered blocks are cut into thin disks and polished on different grades of emery papers to obtain parallel surfaces. The polished surfaces are ultrasonically cleaned to remove dust particles. The polished discs are electroded using sputtered on gold, followed by a thin coating of silver paste (dried at 450 $^{o}$C for 2 minutes) to assure good electrical contact. The dielectric response is measured using a Hewlett-Packard 4194A impedance analyser, which can cover a frequency range from 0.1 kHz to 100 kHz. For low temperature measurements, the samples are placed in a Delta Design 9023 test chamber, which is operated between -160 $^{o}$C and +160 $^{o}$C. The temperature is measured using a Eurotherm temperature controller via a K-type thermocouple mounted directly on the ground electrode of the sample fixture. The analyser, test chamber and Eurotherm temperature controller are interfaced with computer to collect data at 20 different frequencies while cooling at a rate of 2 $^{o}$C/min. Field induced polarization measurements are carried out at 50 Hz using Precision workstation of Radiant Technology, USA. For low temperature PE loop measurement, ceramic sample is kept in cold finger set up submerged in liquid nitrogen and temperature of the sample is measured using RTD sensor mounted directly on the ground electrode of the sample fixture. All the HP-sintered ceramics are tested up to maximum 20 kV/cm external field applied at 50 Hz. Application of more than 20 kV/cm at lower frequencies are not carried out due to sample fixture limitation.

## 3. Results and discussion

### 3.1. Phase analysis using XRD, EDXS and SEM

Figure 1 compares room temperature XRD pattern of sintered PMN ceramics with Gd-substituted PMN (PGMN4-VA/VB/R) ceramics. It is observed that single perovskite phase is present for PMN but minor secondary phase along with pure perovskite phase are present in PGMN4-VA/VB/R. All the major diffraction peaks in the XRD pattern are indexed with perovskite phase of the PMN (JCPDF 22-1199) and other minor peaks are assigned to the pyrochlore phase (JCPDF 37-0071), marked by '*'), GdNbO$_4$ phase (JCPDF 22-1104, marked by 'o'), and MgO (JCPDF 89-4248, marked by an arrow), which are depicted in the inset of Fig. 1. The inset of Fig. 1, clearly reveals formation of the pyrochlore phase for PGMN4-VA in which the charge imbalance is compensated by creating A-site vacancies. No diffraction peak corresponding to the pyrochlore phase is noticed for PGMN4-VB and PGMN4-R. Few minor peaks (marked by 'o') are related to another secondary GdNbO$_4$ phase, whose concentration is found to be less than 1 %.

Fractions of major and minor phases are determined by multiphase Rietveld refinement of the XRD pattern of PGMN4-VA/VB/R ceramics using Fullprof software. For structural analysis, the crystal symmetry Pm3m, Fm3m, Fd3m and I2/a are considered for perovskite, pyrochlore, MgO and GdNbO$_4$ phase, respectively [26]. Figure 2(a-d) depicts fitting of the simulated patterns with the experimental data of PMN and PGMN4-VA/VB/R ceramics. Reliability factor ($R_p$, $R_{wp}$, and $\chi^2$), as mentioned in fig.2a-d, also confirm a good fitting. The phase fractions and lattice parameter are tabulated in Table 1. Segregation of the GdNbO$_4$ phase in PGMN4-VB and PGMN4-R ceramics agrees well with the earlier proposed solubility limit (~2-3%) of Gd in PMN [26,31,33]. Further, the lattice parameter is decreased from 4.0439(4) Å to 4.0419(3) with 4 at.% Gd-substitution, which is consistent with smaller ionic radii of Gd$^{3+}$ for 12 co-ordination ($r_{Gd3+}$ ~1.276 Å) than Pb$^{2+}$ ($r_{Pb2+}$ ~1.49 Å). The lattice parameter is found invariant to the charge compensation mechanism used for Gd-ion substitution at Pb-ion site in PMN.

Enlarged view of diffraction peak within $2\theta$ range from 15 º to 27 º (left panel of the inset of Fig. 1) reveal a noticeable peak for PGMN4-VB and PGMN4-R specimen, which has been assigned to the super-lattice reflections resulting from the CORs induced doubling of unit cell. It is widely known that the 1:1 non-stoichiometric CORs are coexisted with the PNRs in PMN. Size of the CORs (2-5 nm) in PMN is reported to remain unaffected in temperature range 5 to 800 K [24]. Recently, Gd-ion substitution in PMN has shown an enhancement in size of the CORs, which has been related to an increase in the intensity of the super-lattice reflection spots at <½ ½ ½> along <111> unit axis direction. No peak corresponding to the super-lattice has been observed for pure PMN and for PGMN4-VA. Absence of the super-lattice diffraction peak in pure PMN is consistent with earlier reports [34]. However, absence of the super-lattice diffraction peak in the PGMN4-VA may be related to precipitation of the pyrochlore phase. It is known that non-stoichiometric ordering of Mg$^{2+}$ and Nb$^{5+}$ ions in the CORs is a charged state, which is believed to be encapsulated by appositively charged state limiting its size to 2-5 nm. Alio-valent doping in PMN disrupts the charge neutrality, which leads to either precipitation of the pyrochlore phase (Nb-rich) or enhancement of the CORs region (Mg-rich). Creation of the A-site vacancy for PGMN4-A causes precipitation of the pyrochlore phase as a natural way for the charge imbalance compensation mechanism without influencing the CORs size and hence absence of the diffraction peak near ~19º 2θ angle. Enhancement in intensity of the super-lattice peak is consistent with earlier report of alio-valent doped La-PMN, Nd-PMN, Sm-PMN and Pr-PMN etc. ceramics in which the charge imbalance is compensated with changing B-site cation ratio [34-37].

Figure 3(a-d) compares scanning electron microstructure (SEM) images of the fractured surface of PMN, PGMN4-VA, PGMN4-VB, and PGMN4-R, respectively. Both inter-granular and intra-granular fractures along with negligible porosity indicating good sintered density are depicted. Linear intercept method has revealed an average grain size ~ 11 μm for PMN and different average grain size ~ 2.5-12 μm for Gd-substituted PMN. The least average grain size, 2.5 μm is noticed for PGMN4-VA. The average grain size for PGMN4-VB, 4.8 μm is found smaller than 12 μm for PGMN4-R (Table 1), which can be explained by defect inhibited grain growth. In PGMN4-VA, the charge imbalance is compensated by creating Pb-ion vacancies. It may be noticed that for every $2Gd^{3+}$-ion substitution one $Pb^{2+}$ ion vacancy is created, whereas one B-site ion vacancy is created for every 4 $Gd^{3+}$-ion substitution in PGMN4-VB. Larger concentration of the Pb-vacancies in PGMN4-VA compared to that in the B-site vacancies in PGMN4-VB has resulted reduction in grain growth. No-vacancies are created for PGMN-R and hence the grain growth during sintering is not influenced. Similar grain size variations have been reported for rare-earths doped PMN and PMN-PT systems with similar doping level [38-40].

Apart from well-developed large grains of the perovskite phase, few spherical grains (< 1 μm), as indicated by an arrow in Figs. 3(b-d), are also present for PGMN4-VA/VB/R ceramics. Figure 4(a-d) depicts the Pb, Gd, Mg, and Nb elemental mapping on the fractured surface of PMN and PGMN4-VA/VB/R ceramics. Figure 4(a) depicts uniform distribution of the Pb, Mg and Nb elements in the grains of PMN. Segregation of the Gd-ions and Mg-ions is noticed for PGMN4-VA/VB/R ceramics [Fig. 4(b-d)]. Segregation of the Gd-ion is due to formation of the $GdNbO_4$ phase and is consistent with the XRD-analysis. Formation of the $GdNbO_4$ phase has been reported [31,33] to result precipitation of the MgO phase, which is also confirmed in the elemental mapping for PGMN4-VA/VB/R ceramics. It may be noticed that segregation of the Gd-ion in the form of the $GdNbO_4$ phase is not observed in the XRD spectra of PGMN4-VA, which is due to masking of the diffracted peak related to the $GdNbO_4$ phase by the diffracted peak related to the pyrochlore phase. Another interesting observation is large segregation of the Mg-ion for PGMN4-VA compared to PGMN4-VB and PGMN4-R ceramics. It is also noticed that Mg-ion rich and Gd-ion rich regions are close to each other, which strengthen the proposed reaction [26,31] about formation of the GdNbO$_4$ phase from reaction of the Gd$_2$O$_3$ with the Nb-rich pyrochlore phase resulting in the MgO segregation. The detailed microstructural investigation demonstrates an important role of the charge imbalance compensation mechanism on the secondary pyrochlore and GdNbO$_4$ phase in Gd-substituted PMN. This study also reveals non-dependence of the GdNbO$_4$ phase on the charge compensation mechanism. Further, temperature dependent dielectric and ferroelectric measurements are carried to study the influence of charge compensation mechanism on electrical properties.

### 3.2. Dielectric and ferroelectric properties

Temperature dependent dielectric properties ($\varepsilon'$ and $Tan\,\delta$) at different frequencies are performed to investigate effect of the charge compensation mechanism on the relaxor dielectric behaviour. Figure 5(a-d) shows temperature dependent of the dielectric constant and loss tangent at selected frequencies for PMN and PGMN4-VA/VB/R ceramics, respectively. Typical relaxor like dielectric characteristics i.e., broad dielectric maxima ($\varepsilon_m$) with strong frequency dispersion near $T_m$ (T of $\varepsilon_m$ peak) i.e. $T_m$ shifts progressively towards higher temperature with increasing frequency have been observed. This indicates that relaxation process of the PNRs dynamic occurs at multiple time scale. The temperature dependence of $\varepsilon'$ and $Tan\,\delta$ at 1 kHz for PMN and PGMN4-VA/VB/R are compared in Fig. 6(a) and the $\varepsilon_m$ and $T_m$ are presented in Table 2. The $\varepsilon_m \sim 21762$ and $T_m \sim 265$ K for PMN are in good agreement with earlier reports [2,6,26]. The 4 at% Gd-ion substitution in PMN has reduced the $T_m$ ($\sim 30$ K) and $\varepsilon_m$ ($\sim 7600$) and is consistent with increase in size of the CORs, decrease in number or/and sizes of the PNRs and presence of the secondary pyrochlore and GdNbO$_4$ phases leading to reduction in correlation among the PNRs [26,33], which is also in good agreement with earlier reports on A-site substitution by alio-valent ions [37]. The least value of the $\varepsilon_m \sim 4360$ is observed for PGMN4-VA compared to $\sim 7600$ for PGMN4-VB/R, which is due to presence of $\sim 48\%$ pyrochlore phase. It is also observed that PGMN4-VA shows higher value of the $T_m \sim 247$ K compared to 235 K for PGMN4-VB/R, which may be due to low Gd-ion substitution at Pb-site in perovskite lattice and because of the precipitation of GdNbO$_4$ from reaction between the pyrochlore and Gd$_2$O$_3$ [26,31].

Relaxor dielectric characteristics are generally demonstrated by degree of diffuseness ($\delta_A$), degree of relaxation ($\Delta T_m$) and correlation length among the polar nano-regions.

Interacting polar nano-regions is known to grow below the $T_B$ and enhance their correlation length during cooling. Depending upon the correlations among the PNRs, slowing down of polarization fluctuation at $T < T_m$ into randomly orientated polar domains also called super dipolar glass state [41] has been reported. Figure 6(b) compares the $\varepsilon'(T)$ curves at 1 kHz frequency in reduced form $\varepsilon'/\varepsilon'_m$ vs $T/T_m$ for the PMN and PGMN4-VA/VB/R ceramics. Degree of diffusion is observed to increase with Gd-ion substitution and also relate with the defect concentration. The broadness around the $0.8*T/T_m$ clearly reveals large degree of diffusion for the PGMN4-VA compared to the PGMN4-VB/R, which is due to the presence of the pyrochlore phase. Larger broadness is observed for the PGMN4-VB compared to the PGMN4-R, which is attributed to large defect concentrations.

Modified Curie-Weiss relation (Eq. 1) has been used to fit the $\varepsilon'(T)$ above $T_m$ [42]

$$\frac{\varepsilon_A(\omega)}{\varepsilon'(T,\omega)} = 1 + \frac{(T-T_A(\omega))^2}{2\delta_A^2} \tag{1}$$

where, $\varepsilon_A$ ($> \varepsilon_m$), $T_A$ ($< T_m$) and $\delta_A$ (diffuseness) are the fitting parameters, practically independent of frequency and valid for long range of temperatures above the $\varepsilon_m$. Inset of the Fig. 6(c) represents fitting of the $1/\varepsilon_m$ data above $T_m$ to Eq. (1) for PGMN4-VB. The diffuseness parameter $\delta_A$ is presented in Table 2 and also shown in Fig. 6(c) for different charge compensation method. The $\delta_A$ parameter of Gd-substituted PMN is approximately double of that of the PMN, which is attributed to enhancement in the degree of cation's site chemical/charge disorder at the A/B-sites. Further, $\Delta T_m$ (= $T_{m,100kHz}$ - $T_{m,100Hz}$) is related to degree of relaxation and observe to increase from 13.6 (PMN) to 22.4 (PGMN4-R) suggesting enhancement in relaxor characteristics (Table 2).

Frequency dependent dielectric behaviour is a result of the PNRs statistical distribution over wide temperatures range. Number of models are reported to explain frequency dependence of the $T_m$ and the reasonableness of the fitting parameters reveal insight of the interaction among the PNRs [43-46]. Recently, interaction among the PNRs resulting into a critical slowing down of PNRs dynamics at temperature ($T_g$), represented by following relation, is found suitable for Gd-doped PMN [26]. The relation is

$$f = f_o \left(\frac{T}{T_g} - 1\right)^{zv} \tag{2}$$

where, $f_o$ is attempt frequency [$\tau_o = \omega_o^{-1} = (2\pi f_o)^{-1}$ is the microscopic time associated with flipping of fluctuating dipole entities], $zv$ is critical dynamic exponent and $T_g$ is glass transition temperature. Figure 6(d) shows fitting of frequency dependence of the $T_m(f)$ to Eq.

(2) and the corresponding fitted parameters $f_o$, $zv$ and $T_g$ for PMN and PGMN4-VA/VB/R are reported in Table 3 along with the goodness of fit. The open symbols in Fig. 6(d) represent experimental data points and solid line represents the fitted curve. Accurate value of the $T_m$ is determined by fitting the $\varepsilon'(T)$ curve for each frequency in a narrow temperature range around the $T_m$. The fitted parameters $f_o \sim 1.58\pm0.46 \times 10^{15}$ Hz, $T_g = 248.4\pm0.6$ K, and $zv = 10.5\pm0.4$ of PMN are consistent with earlier reported values of PMN [26]. With Gd-substitution, the $f_o$ decreased from $\sim 10^{15}$ Hz to $\sim 10^{13}$ Hz. Recently in PMN, nonergodic ferroelectric cluster glass ground state (also known as "super-dipolar" glass) is reported to develop at a static glass temperature ($T_g = 238$K), which is due to ensemble the PNRs under random electrostatic interaction at high temperature (T > $T_m$) [41]. The fitting parameter $T_g$ is found decreasing from $\sim 224$ K to 193 K and $zv$ is increased from 10.8 to 14.0 with decrease in the defect concentration, which suggests critical slowing down of dynamics of PNRs ensemble (cluster) below this temperature. These fitting parameter are in agreement with the mesoscopic size of the PNRs.

In order to get more insight into the glassy phase, the frequency dependent $\varepsilon'$ (T) data is analysed by an empirical (also known as Mydosh) parameter "$K$" = $\Delta T_{max} / (T_{max} \times \Delta \log (f))$, which has been reported to distinguish between spin glass (SG), cluster glass (CG) and superparamagnetic (SPM) states in magnetic materials. As reported earlier, the related shift $\Delta T_{max}/T_{max}$ is calculated with change of frequency between 100 Hz and ~150 kHz [47]. The larger value of "$K$" > 0.1 is associated to non-interacting spins for SPM system, ~0.03-0.06 for CG weakly interacting spins and smaller values ~0.005-0.01 for SG with strongly interacting spins. Thus, the value of "$K$" is used to distinguish between dipolar glass, super-dipolar glass (CG) and super-paraelectric (SPE) states. The values of "$K$" are presented in Table 3 for PMN and PGMN4-VA/VB/R. A non-ergodic ferroelectric cluster-glass state or "super-dipolar" glass state is expected for PGMN4-VB and PGMN4-R with "$K$" value ~ 0.03. Lower "$K$" value for PGMN4-VA is due to the pyrochlore phase, which might be interfering in the interaction among the PNRs.

Figure 7(a-b) compares *PE*-loop measured at temperature 300 K and 180 K when 20 kV/cm field is switched at 50 Hz frequency for PGMN4-VA/VB/R ceramic samples. At 300 K, the *PE*-loop is "S-shaped" and hysteresis loss free displaying nonlinear nature of the polarization to the applied field and tendency of saturation at large field, as shown in upper side of the inset of Fig. 7(a). The value of maximum polarization, $P_{max} \sim 21.5$ μC/cm$^2$ at 20 kV cm for PMN [shown in the lower side of the inset of Fig. 7(a)] is consistent with the earlier reported result [2,6,26]. Also no change in the *P-E* loop is observed when 20 kV/cm external field is switched at lower frequencies up to 1 Hz. A significant drop in the $P_{max}$ is noticed with 4at% Gd-substitution, which is consistent with earlier study [26]. It may be noticed from the Table 2 that lower $P_{max} \sim 5.6$ μC/cm$^2$ value for PGMN4-VA in comparison to $P_{max} \sim 9.5$ μC/cm$^2$ for PGMN4-VB and PGMN4-R is due to presence of the pyrochlore phase.

Figure 7(b) shows the *P-E* hysteresis loop at 180 K for PGMN4-A/B/R. Inset of the Fig. 7b reveals typical ferroelectric like *P-E* hysteresis loop for PMN, which indicates the development of long range order. The $P_{max}$, $P_r$, and $E_c$ is 35.7 μC/cm$^2$, 30.7 μC/cm$^2$, and 10.5 kV/cm, respectively for PMN is consistent with the earlier report [2,6,26,48]. It is important to note that PMN is able to sustain $P_r$ at 180 K, which confirms the slowing down of polar nano-domains dynamics, but when the applied external field is more than the random field, the nano-polar domains convert into macroscopic domains. The $P_r$, $P_{max}$ and the area under the curve is decreased with 4at% Gd-substitution in PMN, which is due to enhancement in size of the CORs and reduction in correlation among PNRs. The least value of $P_r$ and $P_{max}$ for PGMN4-VA compared to PGMN4-VB/R is due to presence of the pyrochlore phase. This study clearly demonstrates that the charge imbalance created by alio-valent ion substitution at Pb-site in PMN must be carried out by changing B-site cation ratio instead of by creating A-site and B-site vacancies.

## 4. Conclusions

Systematic studies on the charge compensation mechanism concludes that the charge imbalance created by 4 at.% Gd-ion substitution at Pb-site in PMN relaxor ferroelectric must be carried out by changing the B-site cations ratio. Phase analysis confirms the formation of the pyrochlore secondary phase in PGMN4-VA in which charge imbalance is compensated by creating A-site vacancies. Elemental mapping of the fractured surface reveals that all three charge compensation mechanisms are unsuccessful in circumventing minor GdNbO$_4$ phase, which is formed due to the low solubility limit of Gd-ion at Pb-site in perovskite lattice of PMN. Microstructure analysis reveals linear relation between the grain size and the vacancies defect concentration. The lower $P_{max}$ and $\varepsilon_m$ for PGMN4-VA is consistent with the presence of pyrochlore phase. The Mydosh parameter "*K*" reveals critical slowing down of PNRs dynamic resulting ensemble of PNRs to superdipolar glass state for PGMN4-VB/R, whereas the pyrochlore phase in PGMN4-VA causes reduction in interaction among PNRs.

**Acknowledgements**

Authors are grateful to Archana Sagdeo for XRD measurement and D. M. Phase for FE-SEM, EDXS measurements. Mr. Pandey also acknowledges Homi Bhabha National Institute, India for research fellowship.

**References**

[1]  K. Uchino, Relaxor Ferroelectric Devices, Ferroelectrics 151 (1994) 321-330.

[2]  L. E. Cross, Relaxor ferroelectrics, Ferroelectrics 76 (1987) 241-267.

[3]  F. Li, S. Zhang, T. Yang, Z. Xu, N. Zhang, G. Liu, J. Wang, J. Wang, Z. Cheng, Z.-G. Ye, J. Luo, T. R. Shrout, and L.-Q. Chen, The origin of ultrahigh piezoelectricity in relaxor-ferroelectric solid solution crystals, Nat. Commun. 7 (2016) 13807.

[4]  M. J. Krogstad, P. M. Gehring, S. Rosenkranz, R. Osborn, F. Ye, Y. Liu, J. P. C. Ruff, W. Chen, J. M. Wozniak, H. Luo, O. Chmaissem, Z.-G. Ye, and D. Phelan, The relation of local order to material properties in relaxor ferroelectrics, Nat. Mater. 17 (2018) 718-724.

[5]  F. Li, D. Lin, Z. Chen, Z. Cheng, J. Wang, C. Li, Z. Xu, Q. Huang, X. Liao, L.-Q. Chen, T. R. Shrout, and S. Zhang, Ultrahigh piezoelectricity in ferroelectric ceramics by design, Nat. Mater. 17 (2018) 349-354.

[6]  D. Fu, H. Taniguchi, M. Itoh, S. ya Koshihara, N. Yamamoto, and S. Mori, Relaxor $PbMg_{1/3}Nb_{2/3}O_3$: A Ferroelectric with Multiple Inhomogeneities Phys. Rev. Lett. 103 (2009) 207601.

[7]  A. A. Bokov, Z.-G. Ye, Recent progress in relaxor ferroelectrics with perovskite structure, 41(1) (2006) 31-52.

[8]  Z. Xu, S. M. Gupta, D. Viehland, Y. Yan, S. J. Pennycook, Direct Imaging of Atomic Ordering in Undoped and La-Doped $Pb(Mg_{1/3}Nb_{2/3})O_3$, J. Am. Ceram. Soc. 83(1) (2000) 181-188.

[9]  R. A. Cowley, S. N. Gvasaliyac, S. G. Lushnikov, B. Roessli, G. M. Rotaru, Relaxing with relaxors: a review of relaxor ferroelectrics, Adv. Phys. 60 (2011) 229-327.

[10] D. Fu, H. Taniguchi, M. Itoh, S. Mori, Pb(Mg$_{1/3}$Nb$_{2/3}$)O$_3$ (PMN) Relaxor: Dipole Glass or Nano-Domain Ferroelectric, in Advances in Ferroelectrics, edited by A. Pelaiz-Barranco (InTech, 2015).

[11] B. Hehlen, M. Al-Sabbagh, A. Al-Zein, J. Hlinka, Relaxor Ferroelectrics: Back to the Single-Soft-Mode Picture, Phys. Rev. Lett. 117 (2016) 155501.

[12] G. Xu, G. Shirane, J. R. D. Copley, P. M. Gehring, Neutron elastic diffuse scattering study of Pb(Mg$_{1/3}$Nb$_{2/3}$)O$_3$, Phys. Rev. B 69 (2004) 064112.

[13] H. B. Krause, J. M. Cowley, J. Wheatley, Short range ordering in Pb(Mg$_{1/3}$Nb$_{2/3}$)O$_3$, Acta Crystallogr. Sect. A 35 (1979) 1015-1017.

[14] A. A. Bokov, Z.-G. Ye, Dielectric Relaxation in Relaxor Ferroelectrics, J. Adv. Dielectr. 2(2) (2012) 1241010.

[15] R. Blinc, J. Dolinsek, A. Gregorovic, B. Zalar, C. Filipic, Z. Kutnjak, A. Levstik, R. Pirc, Local Polarization Distribution and Edwards-Anderson Order Parameter of Relaxor Ferroelectrics, Phys. Rev. Lett. 83 (1999) 424.

[16] R. Pirc, R. Blinc, Spherical random-bond-random-field model of relaxor ferroelectrics, Phys. Rev. B 60 (1999) 13470.

[17] V. Westphal, W. Kleemann, M. D. Glinchuk, Diffuse phase transitions and random-field-induced domain states of the ''relaxor'' ferroelectric PbMg$_{1/3}$Nb$_{2/3}$O$_3$, Phys. Rev. Lett. 68 (1992) 847.

[18] A. K. Tagantsev and A. E. Glazounov, Mechanism of polarization response in the ergodic phase of a relaxor ferroelectric, Phys. Rev. B 57 (1998) 18.

[19] E. V. Colla, E. Yu. Koroleva, N. M. Okuneva, S.B. Vakhrushev, Long-Time Relaxation of the Dielectric Response in Lead Magnoniobate, Phys. Rev. Lett. 74(9) (1995) 1681-1684.

[20] H. Q. Fan, L. T. Zhanga, L. Y. Zhang, X. Yao, Non-Debye relaxation and the glassy behavior of disordered perovskite ferroelectrics, Solid State Commun. 111 (1999) 541-546.

[21] G. Singh, V. S. Tiwari, V. K. Wadhawan, Crossover from relaxor to normal ferroelectric behaviour in (1-$x$)Pb(Mg$_{1/3}$Nb$_{2/3}$)O$_3$-$x$PbZrO$_3$ ceramic near $x$=0.5, Solid State Commun. 118(8), 407 (2001).

[22] Bidault, E. Husson, P. Gaucher, Substitution effects on the relaxor properties of lead magnesium niobate ceramics, Philos. Mag. B 79(3) (1999) 435-448.

[23] A. D. Hilton, C. A. Randall, D. J. Barber and T. R. Shrout, TEM studies of Pb(Mg$_{1/3}$Nb$_{2/3}$)O$_3$-PbTiO$_3$ ferroelectric relaxors, Ferroelectrics 93 (1989) 379-386.

[24] A. D. Hilton, D. J. Barber, C. A. Randall, T. R. Shrout, Short range ordering in the perovskite lead magnesium niobate, J. Mater. Sci. 25(8) (1990) 3461-3466.

[25] B.-K. Kim, Probing of nanoscaled nonstoichiometric 1:1 ordering in $Pb(Mg_{1/3}Nb_{2/3})O_3$-based relaxor ferroelectrics by Raman spectroscopy, Mater. Sci. Eng. B 94 (2002) 102.

[26] A. H. Pandey, S. M. Gupta, N. P. Lalla, A. K. Nigam, Critical slowing down of polar nano regions ensemble in $Gd^{3+}$-substituted $PbMg_{1/3}Nb_{2/3}O_3$ ceramics, J. of Appl. Phys. 122 (2017) 044101.

[27] M. A. Akbas, P. K. Davies, Thermally Induced Coarsening of the Chemically Ordered Domains in $Pb(Mg_{1/3}Nb_{2/3})O_3$ (PMN)-Based Relaxor Ferroelectrics, J. of Am. Ceram. Soc. 83(1) (2000) 119-123.

[28] G. H. Haertling, Piezoelectric and electrooptic ceramics, in: Relva C. Buchanan (Ed.), Ceramic Materials for Electronics, Processing, Properties and Applications'' in the Series ''Electrical Engineering and Electronics, Marcel Dekker Inc, New York (1986) 139-225.

[29] O. Garcıa-Zaldıvar, A. Pelaiz-Barranco, J. D. S. Guerra, M. E. Mendoza, F. Calderon-Pinar, Influence of the A and B vacancies on the dielectric and structural properties of the PLZT8/60/40 ferroelectric ceramic system, Physica B 406 (2011) 1622–1626.

[30] S. M. Gupta, D. Viehland, Role of charge compensation mechanism in La-modified $Pb(Mg_{1/3}Nb_{2/3})O_3$–$PbTiO_3$ ceramics: Enhanced ordering and pyrochlore formation, J. Appl. Phys. 80 (1996) 5875-5883.

[31] A. H. Pandey, A. K. Srivastava, A. K. Sinha, S. M. Gupta, Investigation of structural, dielectric and ferroelectric properties of Gd-doped lead magnesium niobate ceramics, Mater. Res. Express 2 (2015) 096303.

[32] A. Bhakar, A. H. Pandey, M. N. Singh, A. Upadhyay, A. K. Sinha, S. M. Gupta, T. Ganguli, Structural analysis of lead magnesium niobate using synchrotron powder X-ray diffraction and the Rietveld method, Acta Crystallogr. B 72(3) (2016) 404-409.

[33] A. H. Pandey, V. G. Sathe, S. M. Gupta, Raman spectroscopic investigation of Gd-substituted lead magnesium niobate ceramics $Pb_{1-x}Gd_x(Mg_{1+x/3}Nb_{2-x/3})O_3$ ($0 \leq x \leq 0.1$), J. Alloys Compd. 682, 180 (2016).

[34] D. M. Fanning, I. K. Robinson, S. T. Jung, E. V. Colla, D. D. Viehland, D. A. Payne, Superstructure ordering in lanthanum-doped lead magnesium niobate, J. Appl. Phys. 87(2) (2000) 840-848.

[35] K.-M. Lee, H. M. Jang, and W.-J. Park, Mechanism of 1:1 nonstoichiometric short-range ordering in La-doped Pb(Mg$_{1/3}$Nb$_{2/3}$)O$_3$ relaxor ferroelectrics, J. Mater. Res. 12(6) (1997) 1603-1613.

[36] B.-K. Kim, S.-B. Cha, Synthesis and cationic ordering structure of samarium-doped lead magnesium niobate ceramics, Mat. Res. Bull. 32(6) (1997) 743-747.

[37] B.-K. Kim, S.-B. Cha, J.-W. Jang, Superlattice reflections in Pr$^{3+,4+}$-doped Pb(Mg$_{1/3}$Nb$_{2/3}$)O$_3$, Mater. Lett. 35(1-2) (1998) 1-3.

[38] A. S. Deliormanls, E. Celik, M. Polat, Phase formation and microstructure of Nd$^{+3}$ doped Pb(Mg$_{1/3}$Nb$_{2/3}$)O$_3$ prepared by sol-gel method, J. Mater. Sci.: Mater. Electro 19(6) (2008) 577-583.

[39] J. R. Zhang, Y. C. Zhang, C. J. Lu, W. N. Ye, and J. Su, Effect of La-doping content on the dielectric and ferroelectric properties of 0.88Pb(Mg$_{1/3}$Nb$_{2/3}$)O$_3$-0.12PbTiO$_3$ ceramics, J Mater. Sci.: Mater. Electron. 25 (2014) 653-658.

[40] N. Zhong, P.-H. Xiang, D.-Z. Sun, X.-L. Dong, Effect of rare earth additives on the microstructure and dielectric properties of 0.67Pb(Mg$_{1/3}$Nb$_{2/3}$)O$_3$-0.33PbTiO$_3$ ceramics, Mater. Sci. Eng. B 116(2) (2005) 140-145.

[41] W. Kleeman, J. Dec, Ferroic superglasses: Polar nanoregions in relaxor ferroelectric PMN versus CoFe superspins in a discontinuous multilayer, Phys. Rev. B 94 (2016) 174203.

[42] A. A. Bokov, Y.-H. Bing, W. Chen, Z.-G. Ye, S. A. Bogatina, I. P. Raevski, S. I. Raevskaya, E. V. Sahkar, Empirical scaling of the dielectric permittivity peak in relaxor ferroelectrics, Phys. Rev. B 68 (2003) 052102.

[43] D. Viehland, J. F. Li, S. J. Jang, L. E. Cross, M. Wuttig, Dipolar-glass model for lead magnesium niobate, Phys. Rev. B 43(10) (1991) 8316.

[44] Z.-Y. Cheng, L.-Y. Zhang, and X. Yao, Investigation of glassy behavior of lead magnesium niobate relaxors, J. Appl. Phys. 79(11) (1996) 8615-8619.

[45] S. Kleemann, J. Miga, J. Dec, J. Zhai, Crossover from ferroelectric to relaxor and cluster glass in BaTi$_{1-x}$Zr$_x$O$_3$ ($x$ = 0.25–0.35) studied by non-linear permittivity, Appl. Phys. Lett. 102 (2013) 232907.

[46] J. Souletie, J. L. Tholence, Critical slowing down in spin glasses and other glasses: Fulcher versus power law, Phys. Rev. B 32(1) (1985) 516-519.

[47] J. A. Mydosh, Spin Glasses: An Experimental Introduction (Taylor & Francies, London, 1993) Vol. 125.

[48] X. Zhao, W. Qu, X. Tan, A. A. Bokov, Z.-G. Ye, Electric field-induced phase transitions in (111)-, (110)-, and (100)-oriented Pb(Mg$_{1/3}$Nb$_{2/3}$)O$_3$ single crystals, Phys. Rev. B 75 (2007) 104106.

**Figures and Table Caption**

**Table 1.** Variation of the lattice parameter, different phase concentrations, grain size of PMN and PGMN4-VA/VB/R ceramics.

**Table 2.** Variation of dielectric properties [$\varepsilon_m$, $T_m$, $\delta_A$, $\Delta T_m$ (=$T_{m,100kHz}$ - $T_{m,100Hz}$)], and field induced polarization ($P_{max}$ at 20 kV/cm) of PMN and PGMN4-VA/VB/R ceramics.

**Table 3.** Fitting parameters of critical slowing down glass model [Eq. (3)] to frequency dependent $T_m$ and Mydosh parameter "$K$" for PMN and PGMN4-VA/VB/R ceramics.

**Fig. 1.** Comparison of XRD pattern of the PMN and PGMN4-VA/VB/R ceramics sintered at 1200 $^o$C, where the perovskite phase is indexed by JCPDS 27-1199 and few minor peaks pyrochlore phase are marked by "*"and GdNbO$_4$ marked by "o" MgO is marked by an arrow; inset (a) shows the (½ ½ ½) supperlattice reflection resulting from the doubling of unit cell (Fm3m crystal symmetry), inset (b) is enlarged view in 2$\theta$ range 27$^o$ - 33$^o$ to depict the secondary phases, and inset (c) shows concentration of perovskite phase (%) for PMN and PGMN4-VA/VB/R ceramics.

**Fig. 2.** Determination of lattice parameter and phase fractions using Rietveld refinement fitting of (a) PMN, (b) PGMN-VA, (c) PGMN-VB, and (d) PGMN-R ceramic samples using different phases.

**Fig. 3.** (a) PMN, (b) PGMN-VA, (c) PGMN-VB, and (d) PGMN-R and the spherical grains (size < 1 μm) confirming the second GdNbO$_4$ phase are marked by arrow.

**Fig. 4.** SEM micrograph of the fractured surface of (a) PMN, (b) PGMN-VA, (c) PGMN4-VB, and (d) PGMN4-R ceramic along with Pb, Gd, Mg, and Nb elemental mapping revealing uniform distribution of these ions in the perovskite grains and segregation of the GdNbO$_4$ and MgO phases.

**Fig. 5.** Temperature dependence of the dielectric constant and loss tangent at different frequencies in the 100 Hz to100 kHz range for (a) PMN, (b) PGMN-VA, (c) PGMN-VB, and (d) PGMN-R ceramics.

**Fig. 6.** (a) Comparison of temperature dependence of dielectric constant and loss tangent at 1 kHz frequency for PMN and PGMN4-VA/VB/R ceramics, (b) Comparison of normalized plot $\varepsilon'/\varepsilon'_m$ vs $T/T_m$ of PMN and PGMN4-VA/VB/R ceramics; (c) Variation of diffuseness parameter, $\delta_A$ calculated by fitting $\varepsilon'(T)$ ($f$ = 1 kHz) to Eq. (2) and inset shows fitting of temperature dependence of $1/\varepsilon'(T)$ to Eq. 2 for PGMN4-VB sample, (d) Fitting to critical slowing down glass model, where open symbols represent the temperature of $\varepsilon_m$ ($T_m$) in the frequency range of 0.1-100 kHz and the red dotted line is fitting to Eq. (3).

**Fig. 7.** Comparison of PE-loop recorded at (a) 300 K and (b) 180 K when 20 kV/cm field applied at 50 Hz frequency for PMN and PGMN4-VA/VB/R ceramics.

**Table 1**

| Sample | % Phase deduced from Rietveld refinement | | | | Lattice constant of perovskite phase 'a' (Å) | Average grain size (μm) |
|---|---|---|---|---|---|---|
| | Perovskite | Pyrochlore | MgO | GdNbO$_4$ | | |
| PMN | 100 | - | - | - | 4.0439(4) | 11.0 |
| PGMN4-VA | 58 | 40.82 | 1.19 | - | 4.0418(5) | 2.5 |
| PGMN4-VB | 99.10 | - | - | 0.9 | 4.0419(3) | 4.8 |
| PGMN4-R | 99.18 | - | - | 0.82 | 4.0420(3) | 12.0 |

**Table 2**

| Sample | f = 1 kHz | | $\delta_A$ | $\Delta T_m$ (K) | $P_{max}$ (μC/cm$^2$) E=20 kV/cm |
|---|---|---|---|---|---|
| | $\varepsilon_m$ | $T_m$ (K) | | | |
| PMN | 21762 | 265 | 46 | 13.6 | 21.5 |
| PGMN4-VA | 4360 | 247 | 98 | 17.9 | 5.6 |
| PGMN4-VB | 7668 | 235 | 94 | 20.8 | 9.2 |
| PGMN4-R | 7872 | 235 | 95 | 22.4 | 9.5 |

**Table 3**

| PMN Samples | Critical slowing down model | | | Adj. R-square | Mydosh factor "K" |
|---|---|---|---|---|---|
| | $zv$ | $T_g$ (K) | $f_o$ (Hz) | | |
| PMN | 10.5(±0.4) | 248.4(±0.6) | 1.58(±0.46) x 10$^{15}$ | 0.99995 | 0.017 |
| PGMN4-VA | 10.8(±0.2) | 223.6(±0.4) | 3.23(±0.62) x 10$^{13}$ | 0.99968 | 0.023 |
| PGMN4-VB | 13.5(±0.5) | 199.8(±1.6) | 1.70(±0.32) x 10$^{13}$ | 0.99981 | 0.029 |
| PGMN4-R | 14.0(±0.7) | 193.2(±2.8) | 6.77(±1.04) x 10$^{12}$ | 0.99984 | 0.031 |

Fig. 1

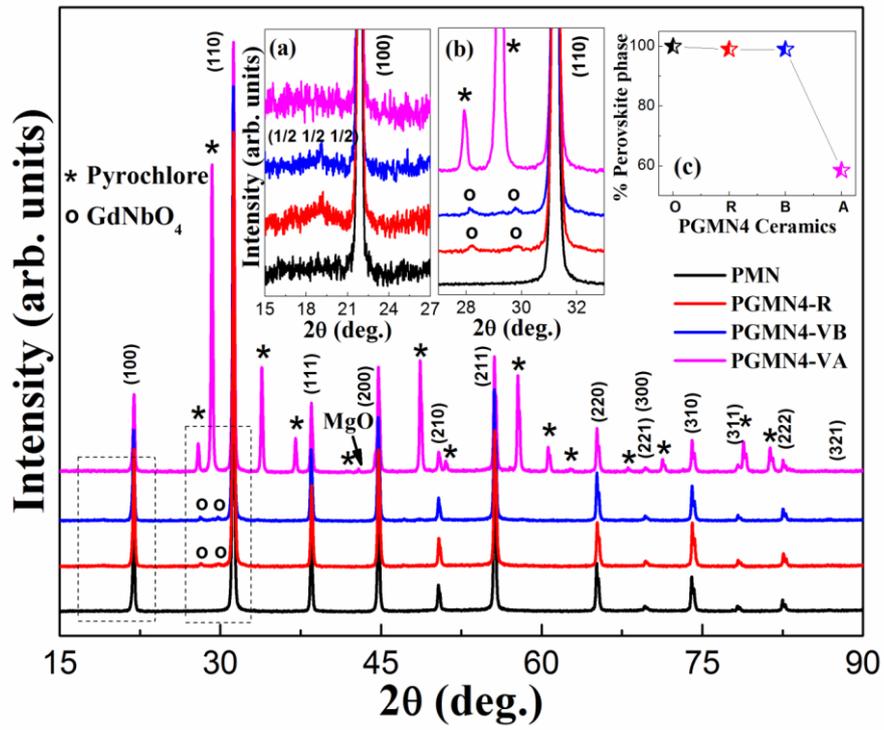

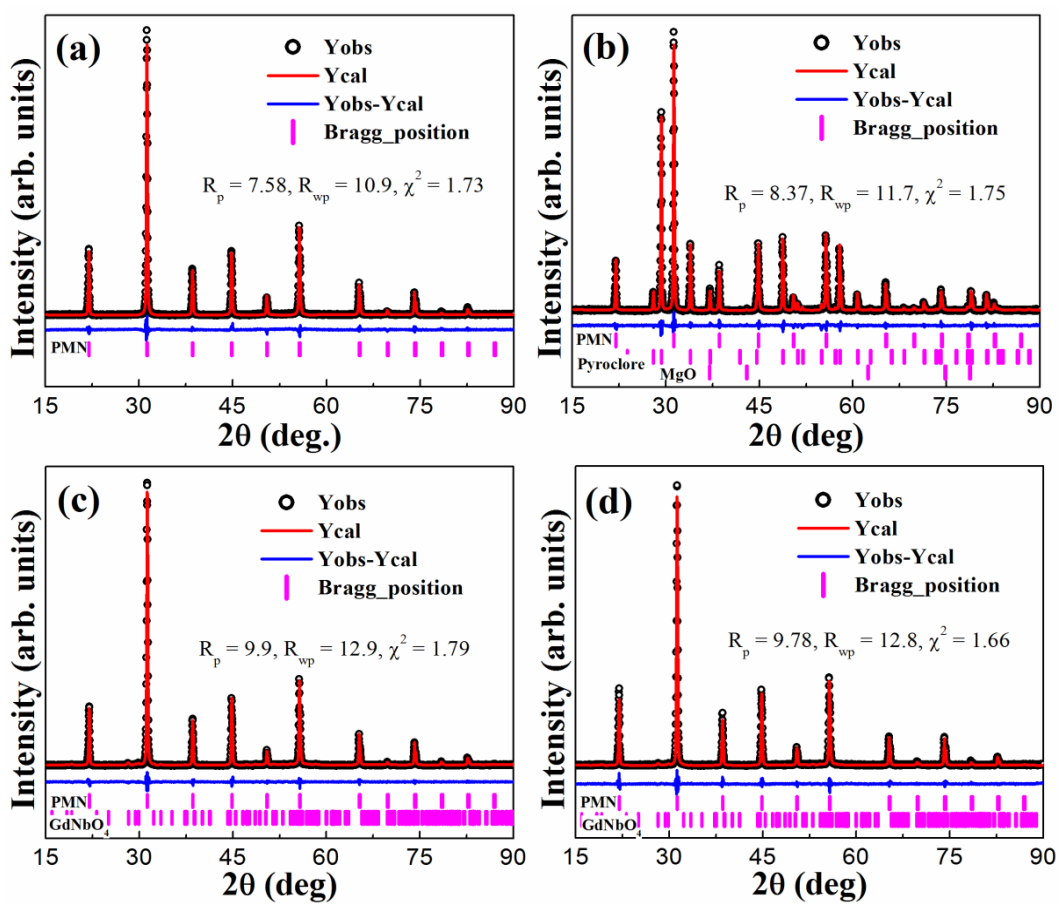

**Fig. 3**

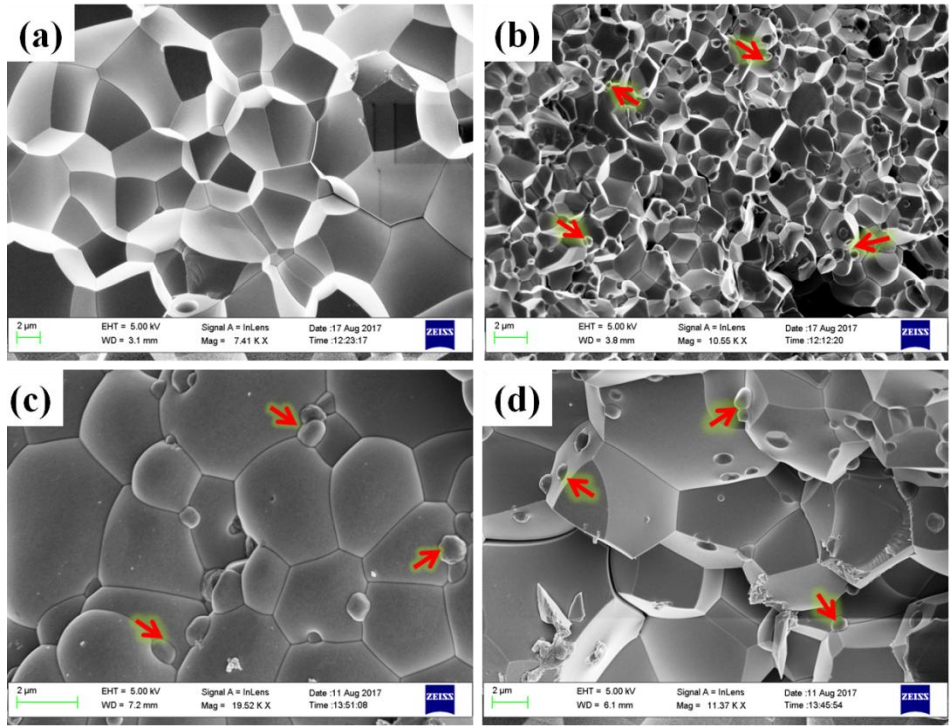

**Fig. 4**

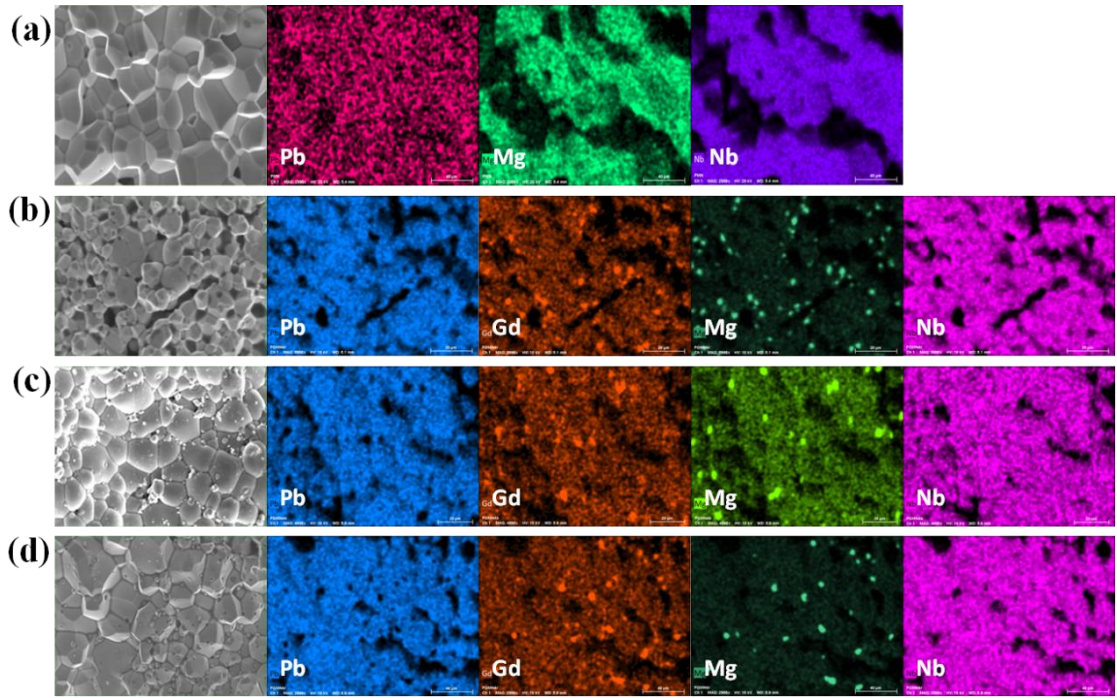

**Fig. 5**

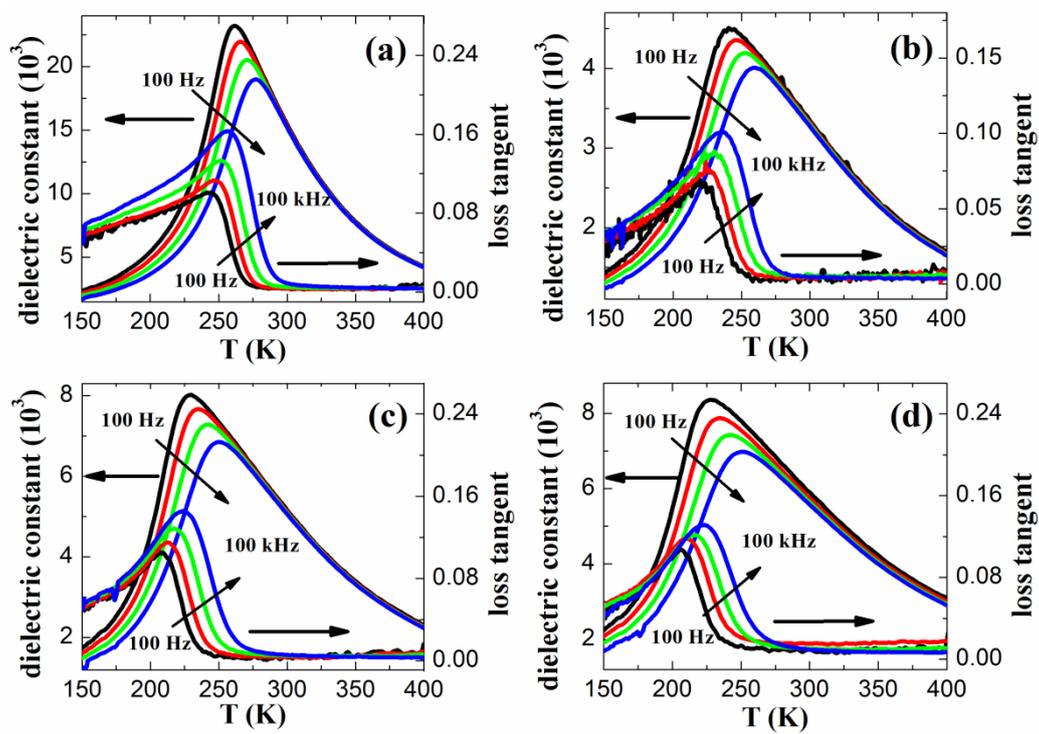

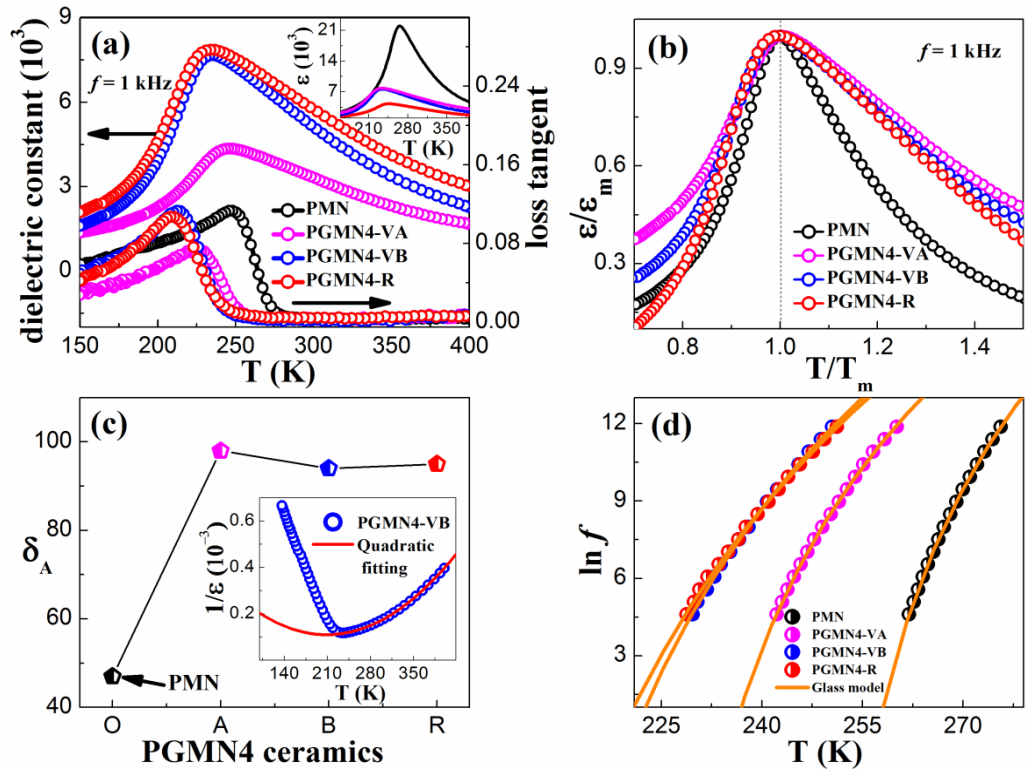

**Fig. 7**

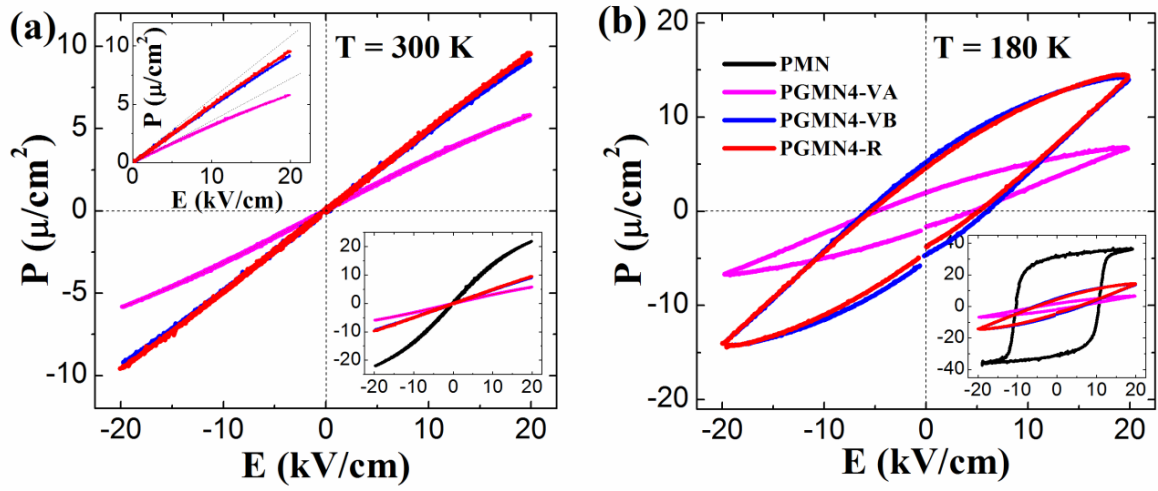